\documentclass[twocolumn,showpacs,amsmath,amssymb,prl]{revtex4}

\usepackage[dvips]{graphicx}
\usepackage{dcolumn}
\usepackage[normalem]{ulem}

\usepackage{amsmath}
\usepackage{amssymb}
\usepackage{upgreek}
\usepackage{psfrag}

\newcommand{\bm}[1]{\boldsymbol{\mathbf{#1}}}
\newcommand{\ud}{\mathrm{d}}
\newcommand{\e}{\bm{E}}
\newcommand{\G}{\bm{G}}
\renewcommand{\S}{\bm{S}}
\newcommand{\bra}{\left\langle}
\newcommand{\ket}{\right\rangle}
\newcommand{\tens}[1]{\bm{#1}}

\newcommand{\im}{\operatorname{Im}}
\newcommand{\tr}{\operatorname{Tr}}

\begin{document}

\title{Near-field interactions and non-universality in speckle patterns \\ produced by a point source in a disordered medium}

\author{A. Caz\'e}
\author{R. Pierrat}
\email{romain.pierrat@espci.fr}
\author{R. Carminati}
\email{remi.carminati@espci.fr}
\affiliation{Institut Langevin, ESPCI ParisTech, CNRS, 10 rue Vauquelin, 75231 Paris Cedex 05, France}

\date{\today}

\begin{abstract}
   A point source in a disordered scattering medium generates a speckle pattern with non-universal features, giving
   rise to the so-called $C_0$ correlation. We analyze theoretically the relationship between the $C_0$ correlation and the statistical 
   fluctuations of the local density of states, based on simple arguments of energy conservation. 
   This derivation leads to a clear physical interpretation of the $C_0$ correlation. Using exact numerical
   simulations, we show that $C_0$ is essentially a correlation resulting from near-field interactions. These interactions
   are responsible for the non-universality of $C_0$, that confers to this correlation a huge potential for sensing and imaging
   at the subwavelength scale in complex media.
    \end{abstract}


\maketitle

% Introduction
Waves propagating in a disordered medium generate a strongly fluctuating
spatial distribution of intensity known as a speckle pattern~\cite{ShengBook}.
The spatial structure of a speckle pattern is often characterized by the
intensity spatial correlation function $\langle I(\bm{r}) I(\bm{r}^\prime) \rangle$, or its
angular counterpart $\langle I(\bm{u}) I(\bm{u}^\prime) \rangle$ where the unit vector $\bm{u}$ defines
an observation direction.
In usual experiments the medium is illuminated by an external beam, and the speckle pattern is observed, e.g., 
in transmission. Short-range and long-range contributions can be identified in the intensity correlation function,
which is written as a sum of three terms denoted by $C_1$ (short range), $C_2$ and $C_3$ (long range)~\cite{Feng1988}. 
These correlations have been widely studied since they are responsible for 
enhanced mesoscopic fluctuations~\cite{AkkermansBook} and their sensitivity to changes
in the medium can be used for imaging in complex media~\cite{SebbahBook}. Moreover,
the possibility of controlling speckle patterns by wavefront shaping has generated new interest~\cite{SLM}.
When the waves are generated by a point source placed inside the medium, a new type of spatial correlation
appear, that has been denoted by $C_0$~\cite{Shapiro1999}. This correlation has features
that make it particularly interesting: It is of infinite range and can dominate the usual long-range correlations,
and it is strongly dependent on the local environment of the source~\cite{Shapiro1999,Skipetrov2000}. 
This non-universality makes $C_0$ a valuable quantity for sensing or imaging in complex media with a high 
sensitivity to the microscopic structure of the medium~\cite{Skipetrov2000}. Moreover, it has been shown that the 
$C_0$ correlation and the fluctuations of the local density of states (LDOS) at the location of the point
source are equal~\cite{Skipetrov2006}. This means that $C_0$ could be obtained from measurements
of LDOS fluctuations instead of an analysis of speckle patterns. In optics, LDOS
fluctuations can be measured from the fluorescence lifetime of
nanoscopic emitters~\cite{Martorell1991,Vallee2006,Vahid2010,Mosk2010,Valentina2010}. They
are expected to provide information on the local environment of the emitter~\cite{Froufe2007,Froufe2008}
and on the photon transport regime~\cite{Mirlin2000,Beenakker2002,Carminati2009,Froufe2009,Pierrat2010}.
%
% Ce travail
In this Letter, we revisit the $C_0$ correlation concept and its connection to LDOS fluctuations in the context of scattering
of electromagnetic waves. We consider electromagnetic waves since the most recent relevant experiments
have been performed in this regime. First, we give a novel derivation of the equality between $C_0$ and
the normalized variance of the LDOS based on energy conservation, showing that the relationship is valid
in any regime, including strong localization~\cite{ShengBook}. 
This derivation leads for the first time to a clear physical interpretation of the $C_0$ correlation.
Second, we analyze the non-universality of $C_0$, and show that it is due to a large extent to near-field interactions. The sensitivity
to the degree of correlation of disorder around the location of the source is demonstrated based on numerical simulations 
on three-dimensional open systems.

% Demonstration analytique
In this first part, we define the $C_0$ correlation and derive its relationship with LDOS fluctuations based on
energy conservation. We consider a non-absorbing disordered system of finite size, embedded in a sphere with radius $ R$ (this
geometry is that used in the subsequent numerical simulations). The system is illuminating by a classical point dipole 
$\bm{p}$ radiating at frequency $\omega$, placed at the center of the cluster at position $\bm{r}_s$. The medium
is assumed statistically isotropic and homogeneous within the sphere of radius $R$. 
For a given configuration of disorder, the LDOS $\rho(\bm{r}_s,\omega)$ can be computed
using the dyadic Green function that describes the response at point $\bm{r}$ to the dipole source
through the relation $\bm{E}(\bm{r})= \mu_0 \, \omega^2 \, \G(\bm{r},\bm{r}_s,\omega) {\bf p}$. 
It is given by $\rho(\bm{r}_s,\omega) = 2 \omega/(\pi c^2) \, \im \left [ \tr  \G(\bm{r}_s,\bm{r}_s,\omega) \right ]$, 
where $\tr$ denotes the trace of a tensor and $c$ is the speed of light in vacuum~\cite{Lagen1996}. 
In terms of the LDOS, the power radiated outside the system,
and averaged over the orientation of the dipole source, reads
$P=\pi \omega^2/(12 \epsilon_0) \, | \bm{p}|^2 \, \rho(\bm{r}_s,\omega) $.
Denoting by $\rho_0$ and $P_0$ the LDOS and the radiated power in vacuum, respectively, we obtain the simple equality
\begin{equation}
   \frac{\rho\left(\bm{r}_s,\omega\right)}{\rho_0\left(\omega\right)}=\frac{P}{P_0}.
   \label{eq:LDOS1}
\end{equation}
As a characterization of the far-field speckle produced by the point source, we consider the angular intensity correlation function
\begin{equation}
   C\left(\bm{u},\bm{u}'\right)=\frac{\bra I\left(\bm{u}\right)I\left(\bm{u}'\right)\ket}{\bra I\left(\bm{u}\right)\ket\bra I\left(\bm{u}'\right)\ket}-1
   \label{eq:correlation}
\end{equation}
where $I\left(\bm{u}\right)$ is the directional radiated power in direction $\bm{u}$, such that $\int_{4\pi}I\left(\bm{u}\right)\ud\bm{u}=P$. 
From Eqs.~(\ref{eq:LDOS1}) and (\ref{eq:correlation}), the fluctuations of the normalized LDOS can be written
\begin{align}\nonumber
   & \bra \frac{\rho^2\left(\bm{r}_s,\omega\right)}{\rho_0^2\left(\omega\right)} \ket = \frac{1}{P_0^2}\iint  \bra I\left(\bm{u}\right)
            I\left(\bm{u}'\right)\ket\ud\bm{u} \, \ud\bm{u}'
\\
   & = \frac{1}{P_0^2}\iint \bra I\left(\bm{u}\right)\ket\bra I\left(\bm{u}'\right)\ket\left[1+C\left(\bm{u},\bm{u}'\right)\right]\ud\bm{u} \, \ud\bm{u}'.
 \label{eq:LDOS2}
\end{align}
Using the hypothesis of statistical isotropy, the averaged directional radiated power reduces to
$\bra I\left(\bm{u}\right)\ket=(P_0/4\pi) \bra \rho\left(\bm{r}_0,\omega\right)\ket/\rho_0\left(\omega\right)$,
and the angular intensity correlation becomes a function of $x = \bm{u}\cdot\bm{u}'$.
This allows us to simplify Eq.~(\ref{eq:LDOS2}) into
\begin{align}\nonumber
   \frac{\bra\rho^2\left(\bm{r}_s,\omega\right)\ket}{\bra\rho\left(\bm{r}_s,\omega\right)\ket^2} & = 1+\frac{1}{16\pi^2}
      \iint C\left(\bm{u}\cdot\bm{u}'\right)\ud\bm{u} \, \ud\bm{u}'
\\
   & = 1+\frac{1}{2}\int_{-1}^{+1} C\left(x\right)\ud x.
   \label{eq:correl2}
\end{align}
The correlation function can be expanded on the basis of Legendre polynomials in the form
$C\left(x\right)=\sum_{n=0}^{\infty}a_nP_n\left(x\right)$. Since $P_0(x)=1$, the first term is constant
and corresponds to an infinite range correlation. To be consistent with initial considerations
of the $C_0$ correlation~\cite{Shapiro1999,Skipetrov2006}, we define the $C_0$ contribution as that given
by the constant term, such that $a_0 = C_0$. The integral in Eq.~(\ref{eq:correl2}) is performed by writing
$C\left(x\right)=\sum_{n=0}^{\infty}a_nP_0\left(x\right)P_n\left(x\right)$ and
using the orthogonality condition of Legendre polynomials $\int_{-1}^{+1}P_n\left(x\right)P_m\left(x\right)\ud x=2\delta_{nm}/\left(2n+1\right)$
where $\delta_{nm}$ is the Kronecker delta. We finally obtain
\begin{equation}
\label{result}
   C_0=\frac{\bra\rho^2\left(\bm{r}_s,\omega\right)\ket}{\bra\rho\left(\bm{r}_s,\omega\right)\ket^2}-1
      =\frac{\operatorname{Var}\left[\rho\left(\bm{r}_s,\omega\right)\right]}{\bra\rho\left(\bm{r}_s,\omega\right)\ket^2}.
\end{equation}
Equation~(\ref{result}) shows that the $C_0$ speckle correlation and the normalized variance of the LDOS at the position of the emitter are the same,
a result that was first derived in Ref.~\onlinecite{Skipetrov2006} based on a diagrammatic approach. Our derivation relies only on considerations
of energy conservation, and is exact within the only assumption of a non-absorbing and statistically isotropic medium. In particular, Eq.~(\ref{result})
holds in all wave transport regimes, from weakly scattering to strongly scattering, including Anderson localization. 
Another feature of our derivation is that it leads to a simple interpretation of the $C_0$ correlation. For a classical point dipole source,
changes in the LDOS correspond to changes in the power transferred to the environment (for a quantum emitter, this corresponds to a change
in the spontaneous decay rate). In a non absorbing medium, this power coincides with the total power radiated in the far field, whose angular
(or spatial) variations generate the speckle pattern. Therefore LDOS fluctuations are transferred into global fluctuations of the speckle patterns,
which themselves are encoded into angular (or spatial) correlations.

\begin{figure}[htbf]
   \centering
   \includegraphics[width=\linewidth]{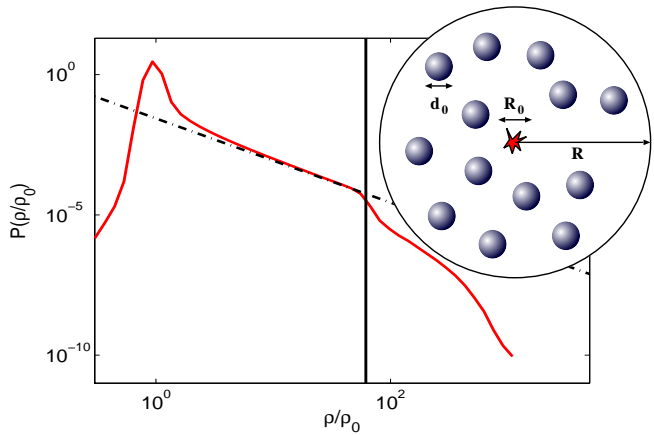}
   \caption{(Color online) Statistical distribution of the normalized LDOS $\rho(\bm{r}_s,\omega)/\rho_0(\omega)$ 
   for a uncorrelated system with correlation parameter (effective volume fraction) $f=0.001\%$ in a double logarithmic plot.
    Others parameters are: $N=100$, $\omega_0=3\times 10^{15}\,\mathrm{Hz}$, $\omega-\omega_0=1\times 10^9\,\mathrm{Hz}$,
     $\gamma=1\times 10^9\,\mathrm{Hz}$,
   $R=1.2 \,\mathrm{\upmu m}$ and $R_0=0.05\,\mathrm{\upmu m}$. The calculation
   are performed with $3\times 10^8$ configurations. This large number of configurations is necessary to correctly describe
   the tail of the distribution. Inset: Schematic view of the system.}
   \label{dist_system}
\end{figure}

% Calculs numeriques en geometrie cluster
We have carried out numerical simulations in order to analyze the dependance of $C_0$ on the local environment of the emitter.
The scattering medium is a three-dimensional cluster of $N$ resonant point scatterers randomly distributed inside a sphere 
with radius $R$. A dipole emitter is placed at the center of the cluster (position $\bm{r}_s$), and is surrounded by a small exclusion 
volume with radius $R_0$. The geometry is shown in the inset in Fig.~\ref{dist_system}.
In the generation of the random configurations of the disorder (i.e. of the positions of the $N$ scatterers), a minimum distance
 $d_0$ is forced between the scatterers. This permits to induce a degree of correlation in the disorder, since this amounts to
 simulating an effective hard-sphere potential between scatterers. One can define an effective volume fraction
$f=N\left(d_0/2\right)^3/\left(R^3-R_0^3\right)$, that can be taken as a measure of the degree of correlation of the disorder
($f$ will be denoted by ``correlation parameter'' in the following).
The scatterers are described by their electric polarizability
$\alpha\left(\omega\right)=-3\pi c^3\gamma/\left[\omega^3\left(\omega-\omega_0+i\gamma/2\right)\right]$ where $\omega_0$ is the resonance frequency,
and $\gamma$ is the linewidth. This corresponds to the polarizability of a resonant non-absorbing point scatterer (which is similar
to that of a two-level atom far from saturation). The scattering cross-section is $\sigma_s(\omega) = (k^4/6\pi) |\alpha(\omega)|^2$,
with $k=\omega/c$. The parameter $k\ell_{\textrm{B}}$ measures the scattering strength, where $\ell_{\textrm{B}}=[\rho\sigma_s(\omega)]^{-1}$
is the independent-scattering (or Boltzmann) mean-free path, $\rho=N/V$ being the density of scatterers.
In all the numerical computations that follows, we have taken
the following set of parameters: $N=100$, $\omega_0=3\times 10^{15}\,\mathrm{Hz}$, $\omega-\omega_0=1\times 10^9\,\mathrm{Hz}$,
$\gamma=1\times 10^9\,\mathrm{Hz}$, $R=1.2 \,\mathrm{\upmu m}$ and $R_0=0.05\,\mathrm{\upmu m}$.
The Boltzmann scattering mean-free path is $\ell_{\textrm{B}}=1.9\,\mathrm{\upmu m}$ so that
$k\ell_{\textrm{B}}=19$ and $R/\ell_{\textrm{B}}=0.63$.

The calculation of the statistical distribution of  the LDOS, from which $C_0$ can be deduced,
amounts to calculating the Green function $\G(\bm{r},\bm{r}_s,\omega)=\G_0(\bm{r},\bm{r}_s,\omega)+\S(\bm{r},\bm{r}_s,\omega)$
for an ensemble of realizations of the scattering medium. Since the free-space Green function $\G_0$ is known analytically,
we only need to calculate the Green function $\S(\bm{r},\bm{r}_s,\omega_0)$ which corresponds to
the scattered field. To proceed, we perform a coupled-dipole numerical computation. The field exciting
scatterer number $j$ is given by the contribution of the dipole source and of all other scatterers,
leading to a set of $3N$ self-consistent equations~\cite{Lax1952}:
\begin{equation}
   \e_j=\mu_0\omega^2\tens{G}_0\left(\bm{r}_j,\bm{r}_s\right)\bm{p}
   +\alpha\left(\omega\right)k^2\sum_{\substack{k=1\\k\ne j}}^N\tens{G}_0\left(\bm{r}_j,\bm{r}_k\right)\e_k
\end{equation}
where $\bm{r}_j$ is the position of scatterer number $j$ and the dependence of the Green functions on $\omega$ have been omitted.
This linear system is solved numerically for each configuration of the disordered medium. Once the exciting electric field
on each scatterer is known, it is possible to compute the scattered field at the source position $\bm{r}_s$
and deduce the Green dyadic, from which the LDOS $\rho$ is readily obtained.
In this numerical approach, near field and far field dipole-dipole interactions and multiple scattering are taken into account rigorously.

% Figure 1 : Distribution et systeme

% Analyse des differents regimes (Fig 1)
We show in Fig.~\ref{dist_system} the statistical distribution of the normalized LDOS $\rho(\bm{r}_s,\omega)/\rho_0(\omega)$  
in the case of a correlation parameter $f=0.001\,\%$ (minimum interparticle distance $d_0=7.5\,\mathrm{nm}$) which corresponds
to an almost uncorrelated medium. The curve exhibits a broad distribution, with values of $\rho/\rho_0$ 
ranging from~$0.2$ to~$1000$. This corresponds to much larger fluctuations than those observed in the single scattering regime, where
$\rho/\rho_0$ slightly deviates from unity~\cite{Froufe2007}. The analysis of the lineshape allows us to distinguish
three regimes. Firstly, the curve covers a zone corresponding to $\rho/\rho_0 < 1$, which means that some configurations
lead to a reduction of the LDOS compared to that in free space. This effect has been analyzed previously
and is due to collective interactions in the multiple scattering regime~\cite{Pierrat2010}. Recent measurements
of the fluorescence lifetime of emitters at the surface of a volume scattering disordered medium seem to provide
evidence of this regime~\cite{Vahid2010}. Secondly, in the region $\rho/\rho_0 > 1$, a power-law decay is observed,
with a statistical distribution behaving as $P(\rho) \propto \rho^{-3/2}$
(the power law is indicated by the dashed line in Fig.~\ref{dist_system}). As originally discussed in Ref.~\onlinecite{Froufe2008},
this behavior is due to near-field interactions between the emitter and the nearest scatterer, in the regime of dipole-dipole interactions,
and in the absence of absorption. Evidence of this regime, and of its influence on the $C_0$ correlation,
have been reported recently~\cite{Mosk2010}.
Thirdly, in the region $\rho/\rho_0 \gg 1$,
the tail of the distribution deviates from the power law $\rho^{-3/2}$. The onset of the deviation depends on the exclusion
volume with radius $R_0$ that encloses the emitter. Indeed, in the single-scattering limit (the emitter interacts chiefly with the 
nearest scatterer), one would observe a sharp cut-off indicated by the vertical solid line in  Fig.~\ref{dist_system}.
This corresponds to the value of $\rho/\rho_0$ given by the near-field dipole-dipole interaction between the emitter
and one scatterer located at a distance $R_0$ ($\rho/\rho_0=62$ in the present case)~\cite{Froufe2008}. The observation
of a tail beyond this single-scattering cut-off is the evidence of near-field interactions with more than one scatterer.
As we shall see below, this tail also contains the information on the local environment of the emitter, and in particular
on the degree of correlation of disorder.

% Figure 2 : Distributions avec variation du parametre de correlation
\begin{figure}[htbf]
   \centering
   \includegraphics[width=\linewidth]{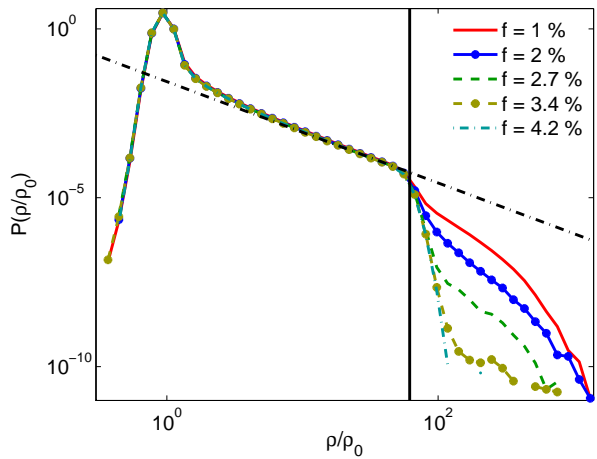}
   \caption{(Color online) Same as Fig.~\ref{dist_system}, with different values of the correlation parameter $f$.
   As a result of near-field interactions, the tail of the distribution is the signature of the local environment of the emitter.}
   \label{dists}
\end{figure}

% Influence de l'ordre local (Figs 2 et 3)
We show in Fig.~\ref{dists} the statistical distribution of $\rho/\rho_0$  for different values of the correlation parameter,
ranging from $1\%$ to $4.2\%$ (i.e. $d_0$ ranging from 111 to 180 nm). 
Increasing $f$ amounts to increasing the level of correlation in the positions of the scatterers
(one imposes an effective hard-sphere potential with a longer range). The tail of the distribution is substantially affected by the level of correlations in the system. Also note that the part of the distribution corresponding to $\rho/\rho_0$ smaller than the single-scatterer
cut-off remains unchanged. This means that the sensitivity of $C_0$ to the local environment of the emitter is driven by the near-field
interactions with the surrounding scatterers, this information being encoded in the tail of the statistical distribution of the LDOS.
Although this tail corresponds to events with a low probability, it is at the core of the $C_0$ correlation concept.

\begin{figure}[htbf]
   \centering
   \includegraphics[width=\linewidth]{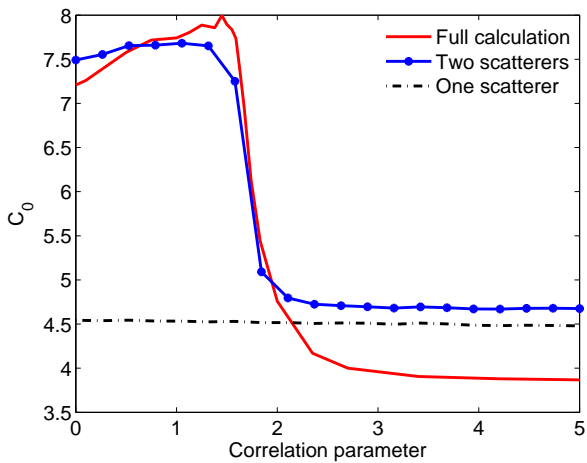}
   \caption{(Color online) $C_0$ speckle correlation versus the correlation parameter $f$. Red solid line: Full numerical simulation
   with the same parameters as in Fig.~\ref{dist_system}. Blue solid line with markers: Calculation considering only the two nearest 
   scatterers (double scattering). Black dashed line: Calculation considering only the nearest scatterer (single scattering).}
   \label{C_0}
\end{figure}

In order to visualize the influence of the correlation of disorder directly on $C_0$, we have plotted in Fig.~\ref{C_0} the values of $C_0$
versus the correlation parameter $f$ obtained from numerical simulations (for the same system as in Figs.~\ref{dist_system} and \ref{dists}). 
A sharp transition is visible at $f\simeq 2 \%$, the value of $C_0$ dropping by a factor of 2.
In order to give a physical interpretation of this behavior, we have also plotted in
Fig.~\ref{C_0} the values of $C_0$ computed by considering the interaction with the nearest scatterer only (black dashed curve),
and with the two nearest scatterers (blue line with markers). 
For large $f$, the emitter essentially interacts with one particle (the red solid line
and the black dashed curve have a similar behavior) and the $C_0$ correlation can be
understood in simple terms. This is the regime found experimentally in Ref.~\onlinecite{Mosk2010}. We stress here that this
regime results from a near-field interaction, so that the value of $C_0$ depends on local microscopic parameters (it cannot be
described with the scattering or transport mean free path as a single parameter). 
For small $f$, the probability of getting more than one scatterer in the vicinity of the emitter becomes non-negligible, and the
behavior of $C_0$ cannot be explained (even qualitatively) with a single-scattering model. 
One sees that by including the interaction with the two nearest scatterers (blue curve with markers), one reproduces nicely
the behavior of the transition. This result demonstrates the high sensitivity of $C_0$ to the level of correlation of the disorder, 
and the key role of near-field interactions. These interactions are responsible for the non-universality of $C_0$.

% Figure 3 : C_0 en fonction du parametre de correlation

% Conclusion
In summary, we have have derived the relation between the $C_0$ speckle correlation
and the LDOS fluctuations using arguments of energy conservation. This simple and exact derivation leads for the first time
to an interpretation of $C_0$ based on the fluctuations of the energy delivered by a classical dipole source to a disordered environment. 
Using exact numerical simulations, we have shown that $C_0$ is essentially a correlation resulting from near-field 
interactions. These interactions give $C_0$ its non-universal character, 
that is reflected in its high sensitivity to the level of correlation of disorder.

This work was supported by the French ANR-06-BLAN-0096 CAROL and by the EU Project {\it Nanomagma} NMP3-SL-2008-214107.

\vspace{-0.5cm}

%%%%%%%%%%%%%%

\end{document}